\documentclass[preprint,aps,nofootinbib]{revtex4}
\usepackage{amsfonts,amsmath,amssymb,amsthm}
\usepackage{latexsym}
\usepackage{bbm,bm}
\usepackage{graphicx}

\newcommand{\ket}[1]{\lvert #1 \rangle}
\newcommand{\bra}[1]{\langle #1 \lvert}
\newcommand{\beq}{\begin{equation}}
\newcommand{\eeq}{\end{equation}}
\newcommand{\beqs}{\begin{eqnarray}}
\newcommand{\eeqs}{\end{eqnarray}}
\begin{document}

\title{Tripartite Entanglement-Dependence of Tripartite Non-locality in Non-inertial Frame}

\author{DaeKil Park$^{1,2}$}

\affiliation{$^1$ Department of Physics, Kyungnam University, Changwon, 631-701, Korea \\
             $^2$ Department of Electronic Engineering, Kyungnam University, Changwon, 631-701, Korea }

\begin{abstract}
The three-tangle-dependence of $S_{max} = \max \langle S \rangle$, where $S$ is Svetlichny operator, are explicitly derived when one party 
moves with an uniform acceleration with respect to other parties in the generalized Greenberger-Horne-Zeilinger and maximally slice
states. The $\pi$-tangle-dependence of $S_{max}$ are also derived implicitly. From the dependence we conjecture that the multipartite 
entanglement is not the only physical resource for quantum mechanical multipartite non-locality.
\end{abstract}

\maketitle

After Einstein-Podolsky-Rosen's seminal paper\cite{epr35} the unusual properties of the quantum correlations became a fundamental issue 
in quantum information theories. This unusual properties become manifest if one examines Bell inequality 
$\langle {\cal B} \rangle \leq 2$ \cite{bell64} by making use of bipartite quantum states. If this inequality is violated, this
fact guarantees the non-locality of quantum mechanics. As Gisin\cite{gisin91} showed, the Bell-type Clauser-Horner-Shimony-Holt (CHSH)\cite{chsh69} 
inequality is violated for all pure entangled two-qubit states. This fact implies that quantum mechanics really exhibits non-local correlations. 
More importantly, the amount of violation $\langle {\cal B} \rangle - 2$ increases when the two-qubit state is entangled more and more.
This fact implies that the origin of the non-local correlations in quantum mechanics is an entanglement of quantum states. 
This remarkable fact can be used to implement the quantum cryptography\cite{ekert91}.

Although the relationship between non-locality and entanglement is manifest to a great extent in two-qubit system, it is 
not straightforward to explore this relationship in multipartite system. Recently, however, understanding in this direction is 
enhanced little bit, especially in three-qubit system. In Ref. \cite{ghose09} the relationship between Svetlichny inequality\cite{svetlichny87}, 
the Bell-type inequality in tripartite system, and tripartite residual entanglement called three-tangle \cite{ckw} was examined 
by making use of the generalized Greenberger-Horne-Zeilinger (GHZ) states $|\psi_g \rangle$ \cite{dur00-1} and the maximally
slice (MS) states $|\psi_s \rangle$ \cite{carteret00} defined as 
\begin{eqnarray}
\label{ghz-class}
& &|\psi_g \rangle = \cos \theta_1 |000\rangle + \sin \theta_1 |111\rangle                             \\   \nonumber
& &|\psi_s\rangle = \frac{1}{\sqrt{2}} \bigg[ |000\rangle + |11\rangle \big\{ \cos \theta_3 |0\rangle + \sin \theta_3 |1\rangle \big\} \bigg].
\end{eqnarray}
The most remarkable fact Re.\cite{ghose09} found is that the $\tau$(three-tangle)-dependence of $S_{max}$, the upper bound of 
expectation value of the Svetlichny operator, for $|\psi_g \rangle$ is 
\begin{eqnarray}
\label{gghz-1}
S_{max} (\psi_g ) = \left\{        \begin{array}{cc}
                           4 \sqrt{1 - \tau}  &  \hspace{1.0cm} \tau \leq 1/3       \\
                           4 \sqrt{2 \tau}    &  \hspace{1.0cm}  \tau \geq 1/3.
                                    \end{array}                       \right.
\end{eqnarray}
Since the Svetlichny inequality is $\langle S \rangle \leq 4$, whose violation guarantees the non-local correlations, Eq. (\ref{gghz-1}) 
shows that $|\psi_g\rangle$ really exhibits non-local correlations in the region $\tau > 1/2$. Unlike two-qubit states, however, 
$S_{max}$ exhibits a decreasing behavior when $\tau \leq 1/3$. This fact strongly suggests that the quantum entanglement is not
the only resource for the multipartite non-locality. It seems to be greatly important issue to find the other resources, which are 
responsible for the non-local properties of quantum mechanics.

%The purpose of this paper is to examine the relationship between tripartite entanglement and $S_{max}$ in non-inertial frame. 
%There are several motivations for our paper. First one is that, needless to say, the non-locality is a fundamental issue in  
%quantum mechanics and quantum information theories. Thus, it is highly interesting to explore the effect of the non-inertial 
%frame on the non-locality in the relativistic setting. Secondly, in order to explore other quantum resources for non-locality we 
%have to derive the relationship between Svetlichny inequality and tripartite entanglement for many quantum states to gain an 
%insight into the relationship. Although the relations are derived for several pure states \cite{ghose09,ajoy10}, still the relationship 
%is not derived for the mixed states. Since analysis in the non-inertial frame naturally generates mixed states, it is possible 
%to extend our experience in the non-locality to the mixtures through this paper. Thirdly, among the mixed states we need a low-rank
%mixed states for the analytical computation of the tripartite entanglement. The reason will be commented later. Since the 
%purification protocol\cite{nielsen00} guarantees that the mixture derived in the analysis of the non-inertial frame is rank-two,
%it is possible to discuss the relationship between Svetlichny inequality and tripartite entanglement on the analytical ground.

The purpose of this paper is to examine the relationship between tripartite entanglement and $S_{max}$ in non-inertial frames.
Since entanglement and non-locality are the two most important concepts in quantum mechanics, the relationship between them is 
at the heart of the foundations of quantum mechanics. Recently, the relations for several $3$-qubit pure states were derived in the
non-relativistic framework\cite{ghose09,ajoy10}. The original purpose of this paper is to extend these relations to the 
relativistic framework. Since, furthermore, the analysis in the non-inertial frames generally transforms a pure state into a mixed 
state due to Unruh decoherence effect\cite{Unruh1,Unruh2}, as a by-product one can derive the relationship between tripartite 
entanglement and $S_{max}$ for various mixed states in this paper.

Although 
similar issue was considered recently in Ref.\cite{smith11}, authors in this reference chose only $\pi$-tangle \cite{ou07} as a tripartite 
entanglement measure. However, the explicit $\pi$-tangle-dependence of $S_{max}$ was not derived in Ref.\cite{smith11}.
As far as we know, furthermore, there are two different tripartite entanglement measures such as three-tangle\cite{ckw}
and $\pi$-tangle\cite{ou07}. Unlike $\pi$-tangle, three-tangle has its own historical background.
In fact, it exactly coincides with the modulus of a Cayley's hyperdeterminant\cite{cay1845,miy03}, which was constructed long ago.
It is also polynomial invariant under the local $SL(2,\mathbb{C})$ transformation\cite{ver03,lei04}. Thus, it seems to be more meaningful 
to derive the three-tangle-dependence of $S_{max}$ explicitly. 

However, the calculation of three-tangle for three-qubit mixed states is much more difficult than that of $\pi$-tangle. Since 
three-tangle for mixed state $\rho$ is defined by convex roof method\cite{benn96,uhlmann99-1}
\begin{equation}
\label{convex-roof}
\tau (\rho) = \min \sum_j P_j \tau (\rho_j),
\end{equation}
where minimum is taken over all possible ensembles of pure states $\rho_j$ with $0 \leq P_j \leq 1$, the explicit computation of three-tangle 
needs to derive an optimal decomposition of the given mixed state $\rho$. It causes difficulties in the analytic computation of the 
three-tangle. Recently, however, various techniques\cite{tangle2,tangle3,tangle4,jung09-1,jung09-2,tangle5} were developed to overcome these difficulties. Still, however, it is highly non-trivial task to compute the three-tangle analytically for high-rank mixed states except very rare cases. 
Fortunately, the mixture derived in this paper is only rank-two. Thus, it is possible to compute the three-tangle analytically using 
various techniques developed in Ref. \cite{tangle2,tangle3,tangle4,jung09-1,jung09-2,tangle5}. 
In this paper we use these techniques to derive the relations between the three-tangle and $S_{max}$ in non-inertial frames.

Now, we assume that Alice, Bob, and Charlie initially share the generalized fermionic GHZ state $|\psi_g \rangle_{ABC}$ or 
the MS state $|\psi_s \rangle_{ABC}$. We also assume that after sharing his own qubit, Charlie moves with respect to Alice and Bob with a uniform acceleration $a$.
Then, Charlie's vacuum and one-particle states $|0\rangle_M$ and $\ket{1}_M$, where the subscript $M$ stands for Minkowski, are transformed
into\cite{alsing06}
\begin{eqnarray}
\label{unruh-1}
& &\ket{0}_M \rightarrow \cos r \ket{0}_I \ket{0}_{II} + \sin r \ket{1}_I \ket{1}_{II}             \\    \nonumber
& &\ket{1}_M \rightarrow \ket{1}_I \ket{0}_{II},
\end{eqnarray}
where the parameter $r$ is defined by 
\begin{equation}
\label{unruh-2}
\cos r = \frac{1}{\sqrt{1 + \exp (-2 \pi \omega c / a)}},
\end{equation}
and $c$ is the speed of light, and $\omega$ is the central frequency of the fermion wave packet\footnote{For bosonic state Eq.(\ref{unruh-1}) is 
changed into
$$ \ket{0}_M \rightarrow \frac{1}{\cosh r} \sum_{n=0}^{\infty} \tanh^n r \ket{n}_I \ket{n}_{II} \hspace{.5cm}
\ket{1}_M \rightarrow \frac{1}{\cosh^2 r} \sum_{n=0}^{\infty} \tanh^n r \sqrt{n+1} \ket{n+1}_I \ket{n}_{II},$$ where
$$ \cosh r = \frac{1}{\sqrt{1 - \exp (-2 \pi \omega c / a)}}.$$}.
Thus, $r=0$ when $a=0$ and $r = \pi/4$ when $a = \infty$.
In Eq. (\ref{unruh-1}) $\ket{n}_I$ and $\ket{n}_{II}$ ($n=0, 1$) are the mode decomposition in the two causally 
disconnected regions in Rindler space. Therefore, Eq. (\ref{unruh-1}) implies that the physical information initially formed 
in region $I$ is leaked into the region $II$, which is a main story of the Unruh effect\cite{Unruh1,Unruh2}. 

Before we discuss on the relationship between Svetlichny inequality and tripartite entanglement, we should comment that
the superselection rule (SSR) of the fermion fields\cite{weinberg} does not allow $|\psi_g \rangle_{ABC}$ and $|\psi_s \rangle_{ABC}$
as fermion states.
This can be easily confirmed by the fact that $|\psi_g \rangle \langle \psi_g|$ and $|\psi_s \rangle \langle \psi_s|$ are not
commute with $(-1)^{\hat{F}} = \mbox{diag} \{1, -1, -1, 1, -1, 1, 1, -1\}$, where $\hat{F}$ is the fermion number operator\cite{ssr1}.
Recently, the SSR and some other subtle issues for the fermion fields were discussed in the context of the relativistic quantum information 
theories\cite{ssr2,ssr3,ssr4}. Furthermore, as discussed in Ref.\cite{ssr1}, this SSR constraint also modifies the definition of the 
three-tangle for the mixed states because the optimal decompositions also should obey the SSR constraint. 
If, therefore, the SSR is taken into account, Eq. (\ref{convex-roof}) yields merely the lower bound of the three-tangle. 

In spite of this fact we will neglect the restriction generated by SSR in this paper. The main reason for this  is that as Weinberg
commented in Ref. \cite{weinberg}, it is always possible to enlarge the symmetry group to a new one that lacks the SSR. Thus, 
it is possible to remove the SSR restriction by extending the symmetry group appropriately. 

%Second reason is that our motivation is to 
%explore the effect of the non-inertial frame on the non-locality simply by making use of Eq. (\ref{unruh-1}). Thus, in order to compare the 
%results derived in this paper to the corresponding results in the inertial frame, we need results on the relation between 
%Svetlichny inequality and tripartite entanglement in the inertial frame. Since these relations for 
%$|\psi_g \rangle_{ABC}$ and $|\psi_s \rangle_{ABC}$ are explicitly given in Ref. \cite{ghose09}, it is convenient to use the 
%results of Ref. \cite{ghose09} for comparison. Thus, our following results might be meaningful in the purely academic aspect, where the map from 
%fermion Fock states to qubit states is performed. The analysis with more realistic physical fields is beyond the scope of this 
%paper.

Using Eq. (\ref{unruh-1}) one can easily show that the Charlie's acceleration makes $|\psi \rangle_{ABC}$ to be 
\begin{equation}
\label{fghz-1}
|\psi \rangle_{ABC} \rightarrow
\big[\cos \theta_1 \cos r \ket{000} + \sin \theta_1 \ket{111} \big]\otimes \ket{0}_{II}
+ \cos \theta_1 \sin r \ket{001} \otimes \ket{1}_{II},
\end{equation}
where $\ket{\alpha\beta\gamma} \equiv \ket{\alpha\beta}_{AB}^M \otimes \ket{\gamma}_I$. Since $\ket{\psi}_{II}$ is a physically 
inaccessible state from region $I$, it is reasonable to take a partial trace over $II$ to average it out. Then, the remaining quantum state 
becomes the following mixed state:
\begin{eqnarray}
\label{fghz-2}
& &\rho_{ABI} = \cos^2 \theta_1 \cos^2 r \ket{000}\bra{000} + \cos^2 \theta_1 \sin^2 r \ket{001}\bra{001} + \sin^2 \theta_1 \ket{111}\bra{111}
                                                                                                             \\   \nonumber
& &\hspace{1.0cm}
+ \sin \theta_1 \cos \theta_1 \cos r \bigg\{\ket{000}\bra{111} + \ket{111}\bra{000} \bigg\}.
\end{eqnarray}
The maximum of the expectation value of the Svetlichny operator, $S_{max}$, for $\rho_{ABI}$ was explicitly derived in Ref.\cite{smith11}, 
and the final expression can be written as 
\begin{equation}
\label{fghz-3}
S_{max} = 4 \max \bigg[ |2 \cos^2 \theta_1 \cos^2 r - 1|, \sqrt{2} |\sin 2 \theta_1 | \cos r \bigg].
\end{equation}
When $a = 0$, Eq. (\ref{fghz-3}) reduces to $S_{max} = 4 \max \left[ |2 \cos^2 \theta_1 - 1|, \sqrt{2} |\sin 2 \theta_1 | \right]$, which 
ensures that the violation of the Svetlichny inequality arises when $\pi/8 < \theta_1 <3 \pi/8$ in a region $0 \leq \theta_1 \leq \pi/2$. 
When $a = \infty$, Eq. (\ref{fghz-3}) reduces to $S_{max} = 4 \max \left[1 - \cos^2 \theta_1, \sin 2 \theta_1 \right]$, which shows that
there is no violation of the Svetlichny inequality.

Now, we discuss on the tripartite entanglement of $\rho_{ABI}$ given in Eq. (\ref{fghz-2}). The computation of its $\pi$-tangle is 
straightforward and the final expression becomes
\begin{equation}
\label{fghz-pi}
\pi_{GGHZ} = \frac{2 + \cos^2 r}{3} \sin^2 2 \theta_1 + \frac{1}{3} \cos^4 \theta_1 \sin^2 2 r.
\end{equation}
When, therefore, $a = 0$, $\pi_{GGHZ}$ becomes $\sin^2 2 \theta_1$, which shows that $\ket{\psi_g}$ is maximally
entangled at $\theta_1 = \pi/4$ and non-entangled at $\theta_1 = 0$ and $\pi/2$. When $a = \infty$, Eq. (\ref{fghz-pi}) reduces to 
$\pi_{GGHZ} = (5/6) \sin^2 2 \theta_1 + (1/3) \cos^4 \theta_1$, which is maximized by $25 / 27 \sim 0.926$ at $\theta_1 = \sin^{-1} (2/3)$
and minimized by zero at $\theta_1 = \pi/2$. The nonvanishing tripartite entanglement at $a \rightarrow \infty$ limit was discussed
in Ref.\cite{hwang11}. This property is a crucial difference from the bosonic bipartite entanglement, which completely vanishes at
$a \rightarrow \infty$ limit\cite{schuller04}.

In order to compute the three-tangle it is convenient to use the spectral decomposition of $\rho_{ABI}$, whose expression is 
\begin{equation}
\label{fghz-tangle-1}
\rho_{ABI} = p \ket{GHZ} \bra{GHZ} + (1-p) \ket{001}\bra{001},
\end{equation}
where $\ket{GHZ} = a \ket{000} + b \ket{111}$ with
\begin{equation}
\label{parameter-1}
p = \cos^2 \theta_1 \cos^2 r + \sin^2 \theta_1    \hspace{.5cm}
a = \frac{\cos \theta_1 \cos r}{\sqrt{\sin^2 \theta_1 + \cos^2 \theta_1 \cos^2 r}}     \hspace{.5cm}
b = \frac{\sin \theta_1}{\sqrt{\sin^2 \theta_1 + \cos^2 \theta_1 \cos^2 r}}.
\end{equation}
In order to derive the optimal decomposition we define
\begin{equation}
\label{optimal-1}
\ket{Z(\phi)} = \sqrt{p} \ket{GHZ} + e^{i \phi} \sqrt{1-p} \ket{001}.
\end{equation}
This has several interesting properties. First, $\rho_{ABI}$ given in Eq.(\ref{fghz-tangle-1}) can be written as 
\begin{equation}
\label{optimal-2}
\rho_{ABI} = \frac{1}{2} \bigg[\ket{Z(\phi)}\bra{Z(\phi)} + \ket{Z(\phi + \pi)}\bra{Z(\phi + \pi)} \bigg].
\end{equation}
Second, the three-tangle of $\ket{Z(\phi)}$ is independent of $\phi$ as $\tau_Z = 4 p^2 a^2 b^2$. If, therefore, 
Eq. (\ref{optimal-2}) is an optimal decomposition, the three-tangle of $\rho_{ABI}$ is also $\tau_{ABI} = 4 p^2 a^2 b^2$.
Since $\tau_{ABI}$ is convex with respect to $p$, this fact guarantees that Eq. (\ref{optimal-2}) is really optimal decomposition
for $\rho_{ABI}$. Using Eq. (\ref{parameter-1}) it is easy to show 
\begin{equation}
\label{t-tangle-1}
\tau_{ABI} = \sin^2 2 \theta_1 \cos^2 r.
\end{equation}
Therefore, combining Eq. (\ref{fghz-3}) and Eq. (\ref{t-tangle-1}) we get the explicit three-tangle-dependence of $S_{max}$ as 
following;
\begin{equation}
\label{depen-1}
S_{max} = 4 \max \bigg[ \sqrt{\cos^2 r - \tau_{ABI}} \cos r - \sin^2 r, \sqrt{2 \tau_{ABI}} \bigg].
\end{equation}
When $a = 0$, it is easy to to show that Eq. (\ref{gghz-1}) is reproduced.

%%%%%%%%%%%%%%%%%%%%%%%%%%%%%%%%%%%%%%%%%%%%%%%%%%%%%%%%%
\begin{figure}[ht!]
\begin{center}
\includegraphics[height=6.5cm]{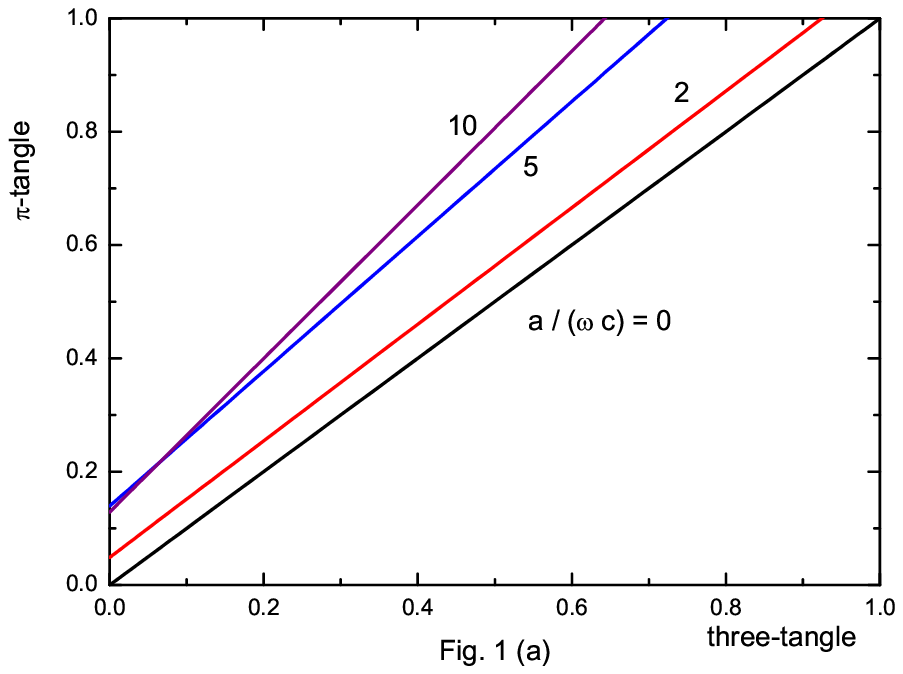}
\includegraphics[height=6.5cm]{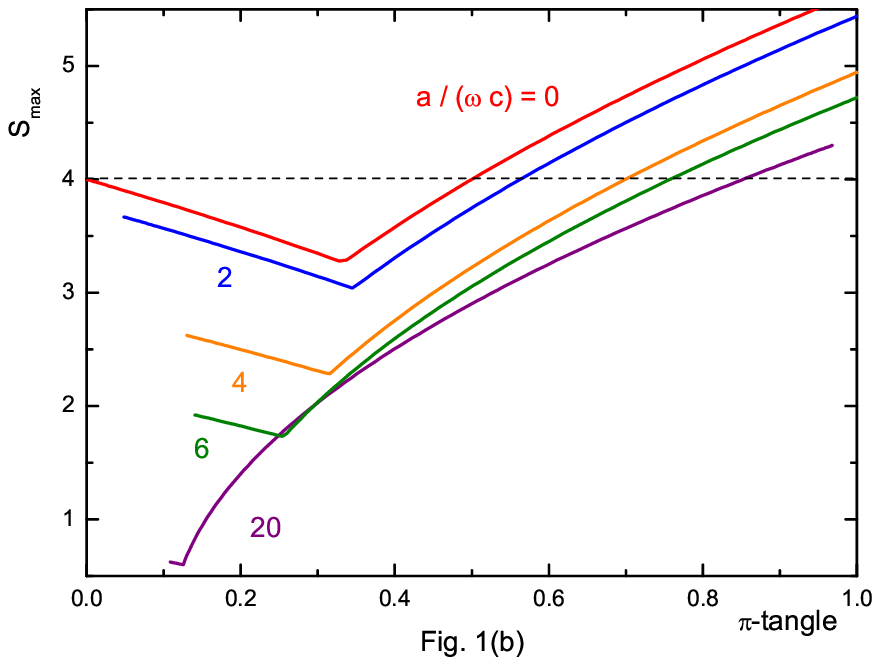}
\includegraphics[height=6.5cm]{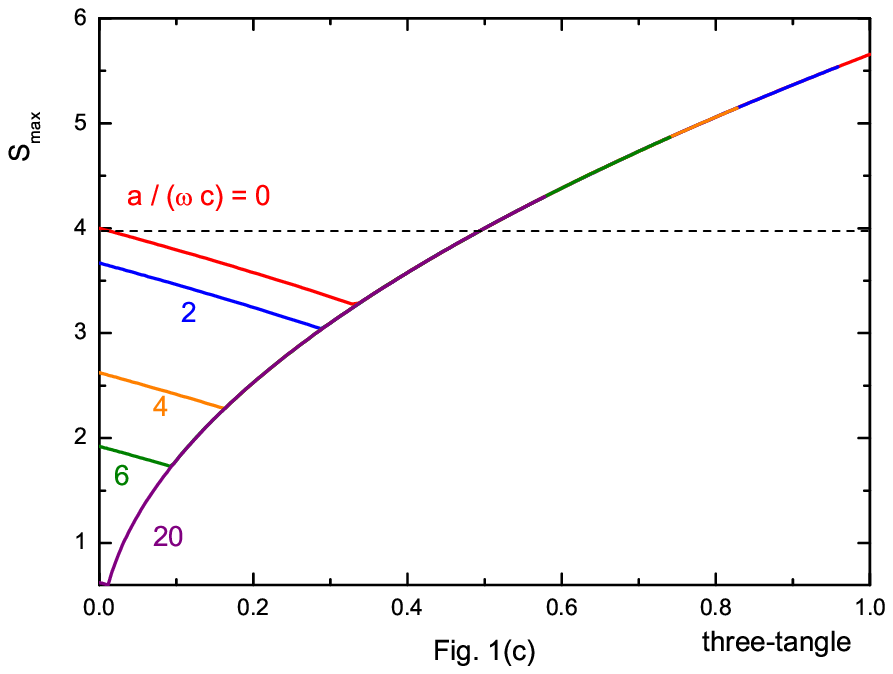}
\caption[fig1]{(Color online) In (a) we plot the $\pi$-tangle (\ref{fghz-pi}) versus three-tangle (\ref{t-tangle-1}). The $\pi$-tangle exhibits 
monotonous behavior with respect to the three-tangle. This fact is plausible because these tangles are two different measures for tripartite 
entanglement. In (b) and (c) we plot the tripartite entanglement-dependence of $S_{max}$. These figures show that $S_{max}$ exhibits a 
decreasing behavior in the small entanglement region. This fact seems to imply that entanglement is not unique physical resource for 
quantum mechanical non-locality.}
\end{center}
\end{figure}
%%%%%%%%%%%%%%%%%%%%%%%%%%%%%%%%%%%%%%%%%%%%%%%%%%%%%%%%%%%

\begin{center}
\begin{tabular}{c||c|c|c|c|c|c|c|c} \hline
$a/\omega c$ & $0$ & $2$ & $4$ & $6$ & $8$ & $10$ & $100$ & $\infty$  \\  \hline \hline
$\pi_*$ & $0.50$ & $0.563$ & $0.70$ & $0.757$ & $0.787$ & $0.806$ & $0.901$ & $1$    \\    \hline
$\tau_*$ & $1$ & $0.959$ & $0.828$ & $0.740$ & $0.687$ & $0.652$ & $0.566$ & $0.5$   \\   \hline
\end{tabular}

\vspace{0.3cm}
Table I: Acceleration dependence of $\pi_*$ and $\tau_*$
\end{center}
\vspace{0.5cm}

In Fig. 1 (a) we plot the three-tangle-dependence of $\pi$-tangle when $a = 0$, $2\omega c$, $5\omega c$, and $10 \omega c$. 
As expected from a fact that these are two different tripartite entanglement measures, $\pi$-tangle is monotonous with respect to 
three-tangle. Fig. 1(a) also shows that regardless of acceleration $a$ $\pi$-tangle is larger than three-tangle, which was 
conjectured in Ref.\cite{ou07,jung09-2}.

Fig. 1(b) and Fig. 1(c) show the tripartite entanglement-dependence of $S_{max}$. As Fig. 1(b) exhibits, the violation of the 
Svetlichny inequality, i.e. $S_{max} > 4$, occurs when $\pi_{ABI} > \pi_*$, where $\pi_*$ increases with increasing $a$. The 
critical value $\pi_*$ is given in Table I for various $a$. As Table I shows, $\pi_*$ approaches $1$ at $a \rightarrow \infty$ limit,
which implies that there is no violation of the Svetlichny inequality in this limit. Fig. 1(c) is a plot for the $\tau_{ABI}$-dependence 
of $S_{max}$ for various $a$. As Fig. 1(c) exhibits, the violation of the Svetlichny inequality occurs when $\tau_{ABI} > 0.5$ for all $a$.
The maximum of the three-tangle, i.e. $\tau_*$, is dependent on Charlie's acceleration $a$. As Table I shows, $\tau_*$ exhibits a decreasing 
behavior with increasing $a$, and eventually approaches $0.5$ in $a \rightarrow \infty$ limit. This fact also indicates that the state shared
initially by Alice, Bob, and Charlie cannot have non-local property in the infinite Charlie's acceleration although it has nonzero tripartite
entanglement.

If Alice, Bob, and Charlie share initially the MS state $\ket{\psi_s}_{ABC}$, Charlie's acceleration changes $\ket{\psi_s}_{ABC}$ into
\begin{eqnarray}
\label{ms-1}
& &\sigma_{ABI}                                                                                                        
= \frac{1}{2} \Bigg[ \cos^2 r \ket{000}\bra{000} + \sin^2 r \ket{001}\bra{001} + \cos^2 \theta_3 \cos^2 r \ket{110}\bra{110}
                                                                                                                      \\   \nonumber
& & \hspace{4.0cm}
                                   + \left( \sin^2 \theta_3 + \cos^2 \theta_3 \sin^2 r \right) \ket{111}\bra{111}
                                                                                                                       \\   \nonumber
& & \hspace{2.0cm}
+ \cos \theta_3 \cos^2 r \left\{ \ket{000}\bra{110} + \ket{110}\bra{000} \right\} 
+ \sin \theta_3 \cos r \left\{ \ket{000}\bra{111} + \ket{111}\bra{000} \right\}                                        \\   \nonumber
& & \hspace{2.0cm}
+ \cos \theta_3 \sin^2 r \left\{ \ket{001}\bra{111} + \ket{111}\bra{001} \right\}
+ \sin \theta_3 \cos \theta_3 \cos r \left\{ \ket{110}\bra{111} + \ket{111}\bra{110} \right\}    \Bigg].
\end{eqnarray}
The maximum of $\langle S \rangle = \mbox{tr} [\sigma_{ABI} S]$ was explicitly computed in Ref.\cite{smith11}, which has 
a form
\begin{equation}
\label{ms-2}
S_{max} = 4 \bigg[\cos^2 \theta_3 \cos^2 2r + 2 \sin^2 \theta_3 \cos^2 r \bigg]^{1/2}.
\end{equation}
Thus, $S_{max} \geq 4$ for $a = 0$ and $S_{max} \leq 4$ for $a = \infty$. 

The $\pi$-tangle for $\sigma_{ABI}$ can be computed straightforwardly and its final expression is 
\begin{equation}
\label{ms-pi-1}
\pi_{MS} = \frac{1}{3} \left[ \sin^2 \theta_3 (2 + \cos^2 r) + \sin^2 r \cos^2 r (1 + \cos^2 \theta_3)^2 \right].
\end{equation}
In order to compute the three-tangle for $\sigma_{ABI}$ we express $\sigma_{ABI}$ in terms of eigenvectors as following:
\begin{equation}
\label{ms-spectral-1}
\sigma_{ABI} = \Lambda_+ \ket{\Psi_+} \bra{\Psi_+} + \Lambda_- \ket{\Psi_-}\bra{\Psi_-}
\end{equation}
where
\begin{eqnarray}
\label{ms-eigen-1}
& &\Lambda_{\pm} = \frac{1 \pm \sqrt{\Delta}}{2}                                                            \\   \nonumber
& &\ket{\Psi_{\pm}} = \frac{1}{{\cal N}_{\pm}} 
\bigg[ X_{\pm} \ket{000} + Y_{\pm} \ket{001} + Z_{\pm} \ket{110} + W_{\pm} \ket{111} \bigg].
\end{eqnarray}
In Eq. (\ref{ms-eigen-1}) $\Delta = \cos^2 \theta_3 + \cos^2 r \left[\sin^2 \theta_3 - \sin^2 r (1 + \cos^2 \theta_3)^2\right]$ and
\begin{eqnarray}
\label{ms-eigen-2}
& &X_{\pm} = \cos r (\mu \pm \sqrt{\Delta}) \hspace{1,0cm} Y_+ = Y_- = \sin \theta_3 \cos \theta_3 \sin^2 r      \\   \nonumber
& &Z_{\pm} = \cos \theta_3 X_{\pm}   \hspace{2.0cm} W_{\pm} = \sin \theta_3 (\cos^2 r \pm \sqrt{\Delta})
\end{eqnarray}
with $\mu = \cos^2 r - \sin^2 r \cos^2 \theta_3$. The normalization constants ${\cal N}_{\pm}$ are 
\begin{eqnarray}
\label{ms-normalization-1}
{\cal N}_{\pm}^2&=& X_{\pm}^2 + Y_{\pm}^2 + Z_{\pm}^2 + W_{\pm}^2                                                \\   \nonumber
&=&\pm 2 \sqrt{\Delta} \left[(1 + \mu) (\cos^2 r \pm \sqrt{\Delta}) - \sin^2 r \cos^2 r \cos^2 \theta_3 (1 + \cos^2 \theta_3)\right].
\end{eqnarray}
Then, it is easy to show $\bra{\Psi_+}\Psi_- \rangle = 0$. Now, we define
\begin{equation}
\label{ms-optimal-1}
\ket{\Phi_{\pm} (\varphi)} = \sqrt{\Lambda_+} \ket{\Psi_+} \pm e^{i \varphi} \sqrt{\Lambda_-} \ket{\Psi_-}.
\end{equation}
Then, $\sigma_{ABI}$ can be written as 
\begin{equation}
\label{ms-optimal-2}
\sigma_{ABI} = \frac{1}{2} \ket{\Phi_{+} (\varphi)}\bra{\Phi_{+} (\varphi)} + 
               \frac{1}{2} \ket{\Phi_{-} (\varphi)}\bra{\Phi_{-} (\varphi)}.
\end{equation}
The three-tangle $\tau(\Phi_{\pm})$ for $\ket{\Phi_{\pm} (\varphi)}$ are 
\begin{equation}
\label{ms-optimal-3}
\tau(\Phi_{\pm}) = 4 | \tilde{X}_{\pm} \tilde{W}_{\pm} - \tilde{Y}_{\pm} \tilde{Z}_{\pm} |^2
\end{equation}
where $\tilde{G}_{\pm} = \sqrt{\Lambda_+} G_+ / {\cal N}_+ \pm e^{i \varphi} \sqrt{\Lambda_-} G_- / {\cal N}_-$ with 
$G = X$, $Y$, $Z$, or $W$. If, thus, Eq. (\ref{ms-optimal-2}) is an optimal decomposition for $\sigma_{ABI}$, the three-tangle becomes
\begin{eqnarray}
\label{ms-optimal-4}
& &\tau(\sigma_{ABI}) = \frac{4 \Lambda_+^2}{{\cal N}_+^4} (X_+ W_+ - Y_+ Z_+)^2 +
                        \frac{4 \Lambda_-^2}{{\cal N}_-^4} (X_- W_- - Y_- Z_-)^2                           \\   \nonumber
& &\hspace{2.0cm}
+ \frac{4 \Lambda_+ \Lambda_-}{{\cal N}_+^2 {\cal N}_-^2} \left\{ (X_+ W_- + X_- W_+) - (Y_+ Z_- + Y_- Z_+) \right\}^2
                                                                                                            \\   \nonumber
& &\hspace{2.0cm}
+ \frac{8 \Lambda_+ \Lambda_-}{{\cal N}_+^2 {\cal N}_-^2} (X_+ W_+ - Y_+ Z_+) (X_- W_- - Y_- Z_-) \cos 2 \varphi.               
\end{eqnarray}
Since $(X_+ W_+ - Y_+ Z_+) (X_- W_- - Y_- Z_-) = \cos^2 r \sin^4 r \cos^4 \theta_3 \sin^6 \theta_3 \geq 0$, we have to choose $\varphi = \pi / 2$
to minimize $\tau (\sigma_{ABI})$. Then, $\tau (\sigma_{ABI})$ simply reduces to 
\begin{equation}
\label{ms-optimal-5}
\tau (\sigma_{ABI}) = \cos^2 r \sin^2 \theta_3.
\end{equation}
It is interesting to note that the three-tangle is much simpler than the $\pi$-tangle. From Eq. (\ref{ms-2}) and Eq. (\ref{ms-optimal-5}) one can 
derive the three-tangle-dependence of $S_{max}$, which is 
\begin{equation}
\label{ms-depen-1}
S_{max} = 4 \sqrt{\cos^2 2r + (5 - 4 \cos^2 r -\tan^2 r) \tau(\sigma_{ABI})}.
\end{equation}
When $a = 0$, Eq. (\ref{ms-depen-1}) reduces to $S_{max} = 4 \sqrt{1 + \tau(\sigma_{ABI})}$. Thus, the violation of the Svetlichny inequality
occurs for all nonzero three-tangle. When $a = \infty$, Eq. (\ref{ms-depen-1}) reduces to $S_{max} = 4 \sqrt{2 \tau(\sigma_{ABI})}$, which 
implies that the violation of the Svetlichny inequality occurs when $\tau(\sigma_{ABI}) > 1 / 2$.

%$\tau(\sigma_{ABI}) \leq 1 / 2$ in the infinite limit.

%%%%%%%%%%%%%%%%%%%%%%%%%%%%%%%%%%%%%%%%%%%%%%%%%%%%%%%%%
\begin{figure}[ht!]
\begin{center}
\includegraphics[height=6.5cm]{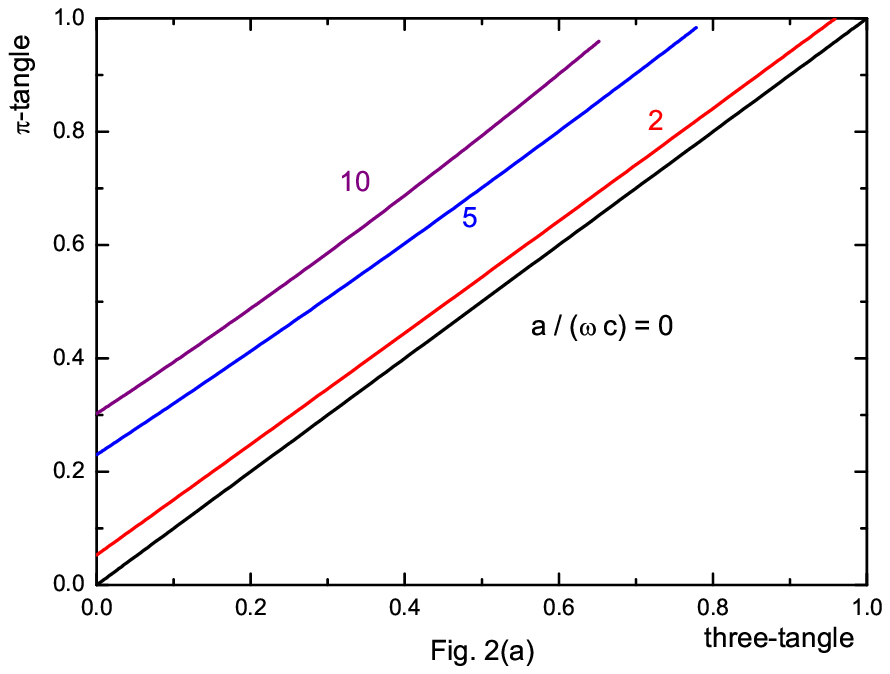}
\includegraphics[height=6.5cm]{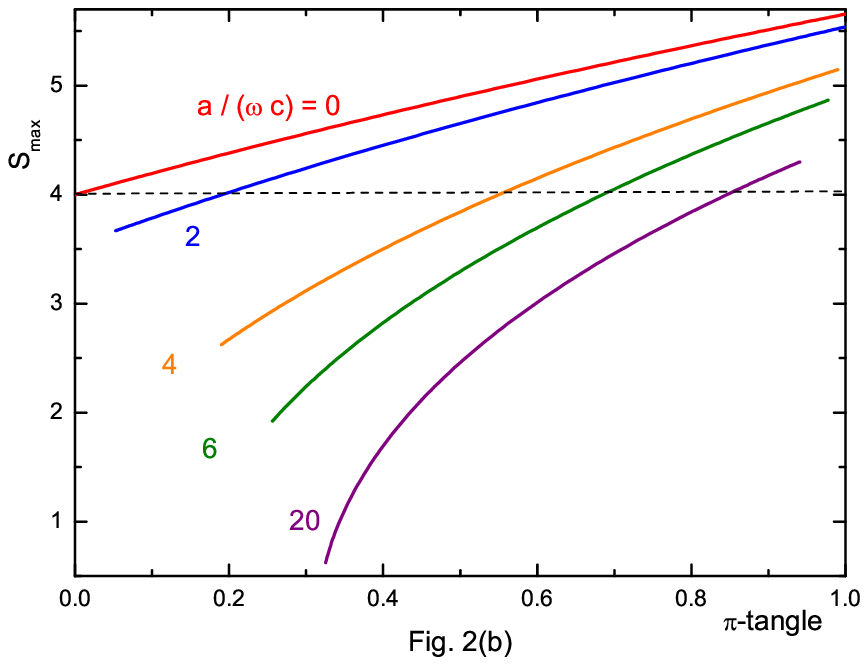}
\includegraphics[height=6.5cm]{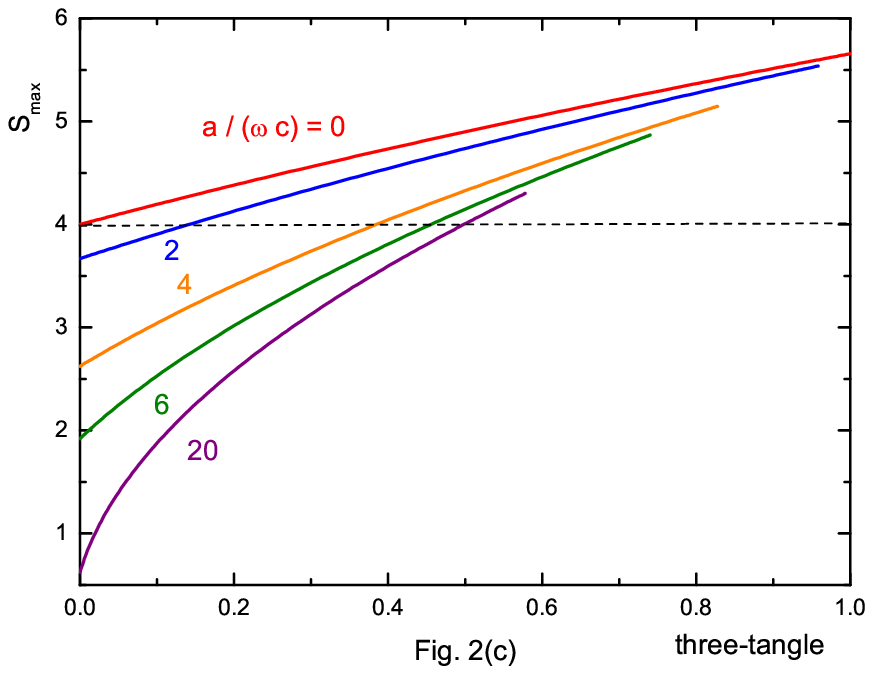}
\caption[fig1]{(Color online) In (a) we plot the $\pi$-tangle (\ref{ms-pi-1}) versus three-tangle (\ref{ms-optimal-5}). As Fig. 1(a) 
the $\pi$-tangle exhibits monotonous behavior with respect to the three-tangle. Regardless of acceleration $a$ the $\pi$-tangle is larger than
the three-tangle, which might be true generally as conjectured in Ref.\cite{ou07,jung09-2}. 
In (b) and (c) we plot the tripartite entanglement-dependence of $S_{max}$. 
Unlike Fig. 1(b) and Fig. 1(c) the decreasing behavior of $S_{max}$ in small entanglement region disappears.}
\end{center}
\end{figure}
%%%%%%%%%%%%%%%%%%%%%%%%%%%%%%%%%%%%%%%%%%%%%%%%%%%%%%%%%%%

In Fig. 2(a) we plot the three-tangle-dependence of $\pi$-tangle for $\sigma_{ABI}$ when $a = 0$, $2\omega c$, $5\omega c$, and $10 \omega c$.
Like Fig. 1(a) the $\pi$-tangle (\ref{ms-pi-1}) is monotonous with respect to the three-tangle (\ref{ms-optimal-5}). Fig. 2(a) also indicates that
$\pi$-tangle is in general larger than three-tangle. In Fig. 2(b) and Fig. 2(c) we plot the tripartite entanglement-dependence of $S_{max}$.
Unlike Fig. 1(b) and Fig. 1(c) there is no decreasing behavior of $S_{max}$ in these figures. From Fig. 2(b) and Fig. 2(c) we know that $\pi_c$
and $\tau_c$ increase with increasing $a$ if the violation of the Svetlichny inequality occurs when $\pi_{MS} > \pi_c$ and 
$\tau (\sigma_{ABI}) > \tau_c$. These critical values are given in Table II for various $a$. Table II shows that $\pi_c \rightarrow 1$ and 
$\tau_c \rightarrow 0.5$ in the infinite acceleration limit. 

\begin{center}
\begin{tabular}{c||c|c|c|c|c|c|c} \hline
$a/\omega c$ & $0$ & $2$ & $4$ & $6$ & $8$ & $10$ & $100$  \\  \hline \hline
$\pi_c$ & $0$ & $0.191$ & $0.250$ & $0.685$ & $0.746$ & $0.780$ & $0.901$    \\    \hline
$\tau_c$ & $0$ & $0.142$ & $0.385$ & $0.456$ & $0.479$ & $0.488$ & $0.5$    \\   \hline
\end{tabular}

\vspace{0.3cm}
Table II: Acceleration dependence of $\pi_c$ and $\tau_c$
\end{center}
\vspace{0.5cm}

If Bob moves, instead of Charlie, with an uniform acceleration, the initial state $\ket{\psi}_{ABC}$ is transformed into
\begin{eqnarray}
\label{ms-bob-1}
& &\sigma_{AIC} = \frac{1}{2} \Bigg[ \cos^2 r \ket{000}\bra{000} + \sin^2 r \ket{010}\bra{010} + \cos^2 \theta_3 \ket{110}\bra{110}
                                     + \sin^2 \theta_3 \ket{111}\bra{111}                                             \\   \nonumber
& &\hspace{2.0cm}
+ \cos r \cos \theta_3 \left\{ \ket{000}\bra{110} + \ket{110}\bra{000} \right\} 
+ \cos r \sin \theta_3 \left\{ \ket{000}\bra{111} + \ket{111}\bra{000} \right\}                                        \\   \nonumber
& &\hspace{4.0cm}
+ \sin \theta_3 \cos \theta_3 \left\{ \ket{110}\bra{111} + \ket{111}\bra{110} \right\}
\Bigg].
\end{eqnarray}
The maximum of $\langle S \rangle = \mbox{tr} [\sigma_{AIC} S]$ was given in Ref. \cite{smith11}, which is 
\begin{equation}
\label{ms-bob-2}
S_{max} = 4 \cos r \left[ \cos^2 \theta_3 + 2 \sin^2 \theta_3 \right]^{1/2}.
\end{equation}
The $\pi$-tangle for $\sigma_{AIC}$ can be straightforwardly computed and the final expression is 
\begin{equation}
\label{ms-bob-3}
\tilde{\pi}_{MS} = \frac{1}{3} \left[1 + \sin^2 \theta_3 - \cos^2 r \cos 2 \theta_3 + \sin^2 r \cos 2r
                   + \sin^2 r \sqrt{\sin^4 r + 4 \cos^2 r \cos^2 \theta_3} \right].
\end{equation}
By similar method one can compute the three-tangle for $\sigma_{AIC}$, which is exactly the same with $\tau (\sigma_{ABI})$ given in 
Eq. (\ref{ms-optimal-5}). Therefore, the three-tangle-dependence of $S_{max}$ in this case is
\begin{equation}
\label{ms-bob-4}
S_{max} = 4 \sqrt{\cos^2 r + \tau(\sigma_{AIC})}.
\end{equation}
Eq. (\ref{ms-bob-4}) implies that the violation of the Svetlichny inequality arises for all nonzero $\tau(\sigma_{AIC})$ when $a = 0$.
It also implies that $\tau(\sigma_{AIC}) \leq 1/2$ when $a \rightarrow \infty$ limit because $S_{max} \leq 4$ in this limit.

In this paper we have examined the tripartite entanglement-dependence of $S_{max} = \max \langle S \rangle$, where $S$ is the Svetlichny
operator, when one party moves with an uniform acceleration $a$ with respect to other parties. If the initial tripartite state is the generalized
GHZ state $\ket{\psi_g}_{ABC}$, the three-tangle-dependence of $S_{max}$ is analytically derived in Eq. (\ref{depen-1}). As Fig. 1 shows,
$S_{max}$ exhibits a decreasing behavior in the small tripartite entanglement region while it exhibits a increasing behavior in the large 
tripartite entanglement region. This fact seems to suggest that the tripartite entanglement is not the only physical resource for the 
tripartite non-locality.
If initial state is the MS state $\ket{\psi_s}_{ABC}$, the explicit relations between $S_{max}$ and three-tangle are 
derived in Eq. (\ref{ms-depen-1}) and Eq. (\ref{ms-bob-4}). In this case the decreasing behavior of $S_{max}$ disappears as Fig. 2 shows. The 
$a$ -dependence of the critical values $\pi_*$, $\tau_*$, $\pi_c$, and $\tau_c$ is summarized in Table I and Table II.

It seems to be interesting to generalize our results to the tripartite bosonic cases\cite{hwang11}. In this case, however, it is highly 
difficult to compute $S_{max}$ in non-inertial frame because the acceleration of one party transforms the qubit system at $a = 0$ into a 
qudit system for nonzero $a$. In order to analyze this issue we should define the Svetlichny-like inequality in the qudit system.

As Eq. (\ref{fghz-3}), Eq. (\ref{ms-2}), and Eq. (\ref{ms-bob-2}) show, the violation of the Svetlichny inequality
does not occur in $a \rightarrow \infty$ limit\cite{friis11} 
even if the tripartite entanglement does not completely vanish in this limit. This fact suggests 
that although there is some connection between the tripartite non-locality and the tripartite entanglement, the entanglement is not
unique resource for the non-locality. Then, what are other physical resources, which are responsible for the non-locality of 
quantum mechanics? As far as we know, we do not have definite answer so far. We will keep on studying this issue in the future.

{\bf Acknowledgement}:
This work was supported by the Kyungnam University
Foundation Grant, 2011.


\begin{thebibliography}{99}
\bibitem{epr35} A. Einstein, B. Podolsky and N. Rosen, {\it Can quantum-mechanical description
of physical reality be considered complete?}, Phys. Rev. {\bf 47} (1935) 777.
\bibitem{bell64} J. S. Bell, {\it On the Einstein-Podolsky-Rosen Paradox}, Physics, {\bf 1}
(1964) 195.
\bibitem{gisin91} N. Gisin, {\it Bell's inequality holds for all non-product states}, 
Phys. Lett. {\bf A 154} (1991) 201.
\bibitem{chsh69} J. F. Clauser, M. A. Horne, A. Shimony and R. A. Holt, {\it Proposed
experiment to test local hidden-variable theories}, Phys. Rev. Lett. {\bf 23} (1969) 880.
\bibitem{ekert91} A. K. Ekert, {\it Quantum cryptography based on Bell’s theorem}, 
Phys. Rev. Lett. {\bf 67} (1991) 661.
\bibitem{ghose09} S. Ghose, N. Sinclair, S. Debnath, P. Rungta, and R. Stock, 
{\it Tripartite Entanglement versus Tripartite Nonlocality in Three-Qubit Greenberger-Horne-Zeilinger-Class States},
Phys. Rev. Lett. {\bf 102} (2009) 250404.
\bibitem{svetlichny87} G. Svetlichny, 
{\it Distinguishing three-body from two-body nonseparability by a Bell-type inequality}, Phys. Rev. {\bf D 35} (1987) 3066.
\bibitem{ckw} V. Coffman, J. Kundu and W. K. Wootters, {\it Distributed entanglement}, Phys. Rev. {\bf A 61} (2000) 052306.
\bibitem{dur00-1} W. D\"{u}r, G. Vidal, and J. I. Cirac, {\it Three qubits can be entangled in two inequivalent ways}, 
Phys. Rev. {\bf A 62} (2000) 062314.
\bibitem{carteret00} H. A. Carteret and A. Sudbery, {\it Local symmetry properties of pure three-qubit states}, J. Phys. {\bf A 33}
(2000) 4981.
\bibitem{ajoy10} A. Ajoy and P. Rungta, {\it Svetlichny's inequlity and genuine tripartite nonlocality in three-qubit pure states},
Phys. Rev. {\bf A 81} (2010) 052334.
%\bibitem{nielsen00} M. A. Nielsen and I. L. Chuang, {\it Quantum Computation and
%Quantum Information} (Cambridge University Press, Cambridge, England, 2000).
\bibitem{Unruh1} W. G. Unruh, {\it Notes on black-hole evaporation}, Phys. Rev. {\bf D 14} (1976) 870.
\bibitem{Unruh2} N. D. Birrel and P. C. W. Davies, {\it Quantum Fields in Curved Space} (Cambridge University Press, Cambridge, England, 1982).
\bibitem{smith11} A. Smith and R. B. Mann, {\it Persistence of Tripartite Nonlocality for Non-inertial observers}, arXiv:1107.4633 [quant-ph].
\bibitem{ou07} Y. U. Ou and H. Fan, {\it Monogamy inequality in terms of negativity for three-qubit states}, Phys. Rev. {\bf A 75} (2007) 062308.
\bibitem{cay1845} A. Cayley, {\it On the Theory of Linear Transformations}, Cambridge Math.
J. {\bf 4} (1845) 193.
\bibitem{miy03} A. Miyake, {\it Classification of multipartite entangled states
by multidimensional determinants}, Phys. Rev. {\bf A 67} (2003) 012108.
\bibitem{ver03} F. Verstraete, J. Dehaene and B. D. Moor, {\it Normal forms and entanglement
measures for multipartite quantum states}, Phys. Rev. {\bf A 68} (2003) 012103.
\bibitem{lei04} M. S. Leifer, N. Linden and A. Winter, {\it Measuring polynomial invariants
of multiparty quantum states}, Phys. Rev. {\bf A 69} (2004) 052304.
\bibitem{benn96} C. H. Bennett, D. P. DiVincenzo, J. A. Smokin and W. K. Wootters,
{\it Mixed-state entanglement and quantum error correction}, Phys. Rev. {\bf A 54}
(1996) 3824.
\bibitem{uhlmann99-1} A. Uhlmann, {\it Fidelity and concurrence of conjugate states},
Phys. Rev. {\bf A 62} (2000) 032307.
\bibitem{tangle2} R. Lohmayer, A. Osterloh, J. Siewert and A. Uhlmann, {\it Entangled
Three-Qubit States without Concurrence and Three-Tangle}, Phys. Rev. Lett. {\bf 97}
(2006) 260502.
\bibitem{tangle3} C. Eltschka, A. Osterloh, J. Siewert and A. Uhlmann, {\it Three-tangle
for mixtures of generalized GHZ and generalized W states}, New J. Phys. {\bf 10} (2008)
043014.
\bibitem{tangle4} E. Jung, M. R. Hwang, D. K. Park and J. W. Son, {\it Three-tangle
for Rank-$3$ Mixed States: Mixture of Greenberger-Horne-Zeilinger, W and flipped W states},
Phys. Rev. {\bf A 79} (2009) 024306.
\bibitem{jung09-1} E. Jung, D. K. Park, and J. W. Son, {\it Three-tangle does not properly
quantify tripartite entanglement for Greenberger-Horne-Zeilinger-type state},
Phys. Rev. {\bf A 80} (2009) 010301(R).
\bibitem{jung09-2} E. Jung, M. R. Hwang, D. K. Park, and S. Tamaryan, {\it Three-Party Entanglement in Tripartite Teleportation
Scheme through Noisy Channels}, Quant. Inf. Comp. {\bf 10} (2010) 0377.
\bibitem{tangle5} G. J. He, X. H. Wang, S. M. Fei, H. X. Sun, and Q. Y. Wen, {\it Three-tangle for high-rank mixed states}, 
Commun. Theor. Phys. {\bf 55} (2011) 251.
\bibitem{alsing06} P. M. Alsing, I. Fuentes-Schuller, R. B. Mann, and T. E. Tessier, {\it Entanglement of Dirac fields in noninertial frames},
Phys. Rev. {\bf A 74} (2006) 032326.
\bibitem{weinberg} S. Weinberg, {The quantum Theory of Fields I} (Cambridge University Press, Cambridge, 1995).
\bibitem{ssr1} P. Caban, K. Podlaski, J. Rembieli\'nski, A. Smoli\'nski, and Z. Walczak, {\it Entanglement and tensor product 
decomposition for two fermions}, J, Phys, A: Math. Gen. {\bf 38} (2005) L79.
\bibitem{ssr2} K. Br\'adler, {\it On two misconceptions in current relativistic quantum information} [arXiv:1108.5553 (quant-ph)].
\bibitem{ssr3} K. Br\'adler and R. J\'auregui, {\it Comment on ``Fermionic entanglement ambiguity in noninertial frames''}, 
Phys. Rev. {\bf A 85} (2012) 016301.
\bibitem{ssr4} M. Montero and E. Mart\'in-Mart\'inez, {\it Reply to ``Comment on `Fermionic entanglement ambiguity in noninertial frames' ''},
Phys Rev. {\bf A 85} (2012) 016302.
\bibitem{hwang11} M. R. Hwang, D. K. Park, and E. Jung, {\it Tripartite entanglement in a noninertial frame}, Phys. Rev. {\bf A 83} (2011)
012111.
\bibitem{schuller04} I. Fuentes-Schuller and R. B. Mann, {\it Alice Falls into a Black Hole: Entanglement in Noninertial Frames},
Phys. Rev. Lett. {\bf 95} (2005) 120404.
\bibitem{friis11} N. Friis, P. K\"ohler, E. Mart\'in-Mart\'inez, and R. A. Bertlmann, {\it Residual entanglement of accelerated fermions 
is not nonlocal}, Phys. Rev. {\bf A 84} (2011) 062111.



%\bibitem{bennett93} C. H. Bennett, G. Brassard, C. Cr\'{e}peau, R. Jozsa,
%A. Peres and W. K. Wootters, Phys. Rev. Lett. {\bf 70} (1993) 1895.
%\bibitem{cryptography} C. H. Bennett and G. Brassard, in Proceedings of the IEEE International Conference on
%Computer, Systems, and Signal Processing, Bangalore, India (IEEE, New York, 1984), pp. 175-179;
%A. K. Ekert, Phys. Rev. Lett. {\bf 67} (1991) 661;
% C. A. Fuchs, N. Gisin, R. B. Griffiths, C. S. Niu and A. Peres, Phys.
%Rev. {\bf A 56} (1997) 1163 [quant-ph/9701039].
%\bibitem{qc} G. Vidal, Phys. Rev. Lett. {\bf 91} (2003) 147902 [quant-ph/0301063];
%M. A. Nielsen and I. L. Chuang, {\it Quantum Computation and
%Quantum Information} (Cambridge University Press, Cambridge, England, 2000).
%\bibitem{horodecki07} R. Horodecki, P. Horodecki, M. Horodecki, and K. Horodecki, 
%Rev. Mod. Phys. {\bf 81} (2009) 865 [quant-ph/0702225] and references therein.
%\bibitem{inertial} M. Czachor, Phys. Rev. {\bf A 55} (1997) 72 [quant-ph/9609022]; 
%P. M. Alsing and G. J. Milburn, Quantum Inf. Comput. {\bf 2} (2002) 487 [quant-ph/0203051]; 
%A. Peres, P. F. Scudo and R. Terno, Phys. Rev. Lett. {\bf 88} 230402 [quant-ph/0203033];
%R. M. Gingrich and C. Adami, Phys. Rev. Lett. {\bf 89} (2002) 270402 [quant-ph/0205179];
%A. Peres and D. R. Terno, Rev. Mod. Phys. {\bf 76} (2004) 93 [quant-ph/0212023].

%\bibitem{hawking} S. W. Hawking, Nature, {\bf 248} (1974) 30; Commun. Math. Phys. {\bf 43} (1975) 199.
%\bibitem{martinez10} E. Mart\'in-Mart\'inez, L. J. Garay and J. Le\'on, Phys. Rev. {\bf D82} (2010) 064006 [arXiv:1006.1394 (quant-ph)].
%\bibitem{teleportation} P. M. Alsing and G. J. Milburn, Phys. Rev. Lett. {\bf 91} (2003) 180404 [quant-ph/0302179]; 
%P. M. Alsing, D. McMahon, and G. J. Milburn, quant-ph/0311096; K. Shiokawa, arXiv:0910.1715 (gr-qc).
%\bibitem{debate} N. B. Narozhny, A. M. Fedotov, B. M. Karnakov, V. D. Mur, and V. A. Belinskii, Phys. Rev. {\bf D 65} (2001) 025004 [hep-th/9906181];
%A. M. Fedotov, N. B. Narozhny, V. D. Mur, and V. A. Belinski, Phys. Lett. {\bf A 305} (2002) 211 [hep-th/0208061]; 
%V. A. Belinski, N. B. Narozhny, A. M. Fedotov, and V. D. Mur, Phys. Lett. {\bf A 331} (2004) 349 [hep-th/0306191].
%\bibitem{refutation} S. A. Fulling and W. G. Unruh, Phys. Rev. {\bf D 70} (2004) 048701; 
%N. B. Narozhny, A. M. Fedotov, B. M. Karnakov, V. D. Mur, and V. A. Belinskii, Phys. Rev. {\bf D 70} (2004) 048702.
%\bibitem{cay1845} A. Cayley, Cambridge Math. J. {\bf 4} (1845) 193; 
%A. Miyake, Phys. Rev. {\bf A67} (2003) 012108 [quant-ph/0206111].
%\bibitem{ver03} F. Verstraete, J. Dehaene and B. D. Moor, Phys. Rev. {\bf A68} (2003) 012103
%[quant-ph/0105090]; M. S. Leifer, N. Linden and A. Winter, Phys. Rev. {\bf A69} (2004) 052304 [quant-ph/0308008].
%\bibitem{green89} D. M. Greenberger, M. Horne, and A. Zeilinger, {\it Bell's Theorem,
%Quantum Theory, and Conceptions of the Universe}, edited by M. Kafatos (Kluwer,
%Dordrecht, 1989) p 69.
%\bibitem{dur00-1} W. D\"{u}r, G. Vidal, and J. I. Cirac, Phys. Rev. {\bf A62} (2000) 062314 [quant-ph/0005115].
%\bibitem{hawking76} S. W. Hawking, Phys. Rev. {\bf D14} (1976) 2460.
%\bibitem{wootters} S. Hill and W. K. Wootters, Phys. Rev. Lett. {\bf 78} (1997) 5022 
%[quant-ph/9703041]; W. K. Wootters, Phys. Rev. Lett.
%{\bf 80} (1998) 2245 [quant-ph/9709029].
%\bibitem{tri-teleportation} A. Karlsson and M. Bourennane, Phys. Rev. {\bf A58} (1998) 4394.









\end{thebibliography}
\end{document}